\documentclass[aps,prb,twocolumn,showpacs]{revtex4}
\usepackage{tabularx,graphicx}
\input epsf
\begin{document}
\title{Simple model of anisotropic pairing with repulsive interactions.}
\author{F. Guinea}
\affiliation{
Instituto de Ciencia de Materiales de Madrid,
CSIC, Cantoblanco, E-28049 Madrid, Spain.}
\date{\today}
\begin{abstract}
A simple tight binding model with repulsive interactions is studied.
The inclusion of more than one orbital per site leads to
assisted hopping effects, and, when the orbitals involved have
different symmetries, to an anisotropic superconducting phase.
Superconductivity exists for all fillings, and for all values of
the on site repulsion.
\end{abstract}
\maketitle
{\it Introduction.}
The existence of
superconductivity in models of strongly correlated electrons with 
only repulsive interactions has attracted a great deal of attention
in recent times. It was established, after the formulation
of the BCS theory, that the metallic state is unstable towards
anisotropic superconductivity, due to the angular dependence of
the dielectric constant\cite{KL65}. This Kohn-Luttinger instability
is greatly enhanced when the Fermi surface in anisotropic\cite{CL92,GGV96,H99}.
By using RPA or model dielectric functions, it can also be shown
that isotropic Fermi surfaces give rise to anisotropic
superconductivity\cite{S95}.
Varied numerical evidence suggests that models with purely
repulsive interactions can lead to anisotropic 
superconductivity\cite{WS99,CS00,LK00,Metal00}.
 
An alternative scheme which leads
to a superconducting 
ground state starting from models with 
repulsive interactions was proposed in\cite{HM89,H93}. This model includes
an assisted hopping term which arises naturally when considering
many orbitals at each lattice site. This assisted hopping term
strongly favors the existence of a
superconducting ground state\cite{HM89,H93,Aetal94}.
In its standard version, this
model leads to isotropic
superconductivity for small hole concentrations, 
and for a moderate value of the onsite repulsion. In this work, we
present an extension of this model which has an anisotropic
superconducting ground state for all dopings and arbitrary values of
the onsite repulsion.

\begin{figure}
 \begin{center}
        \leavevmode
        \epsfxsize=85mm 
        \epsfbox{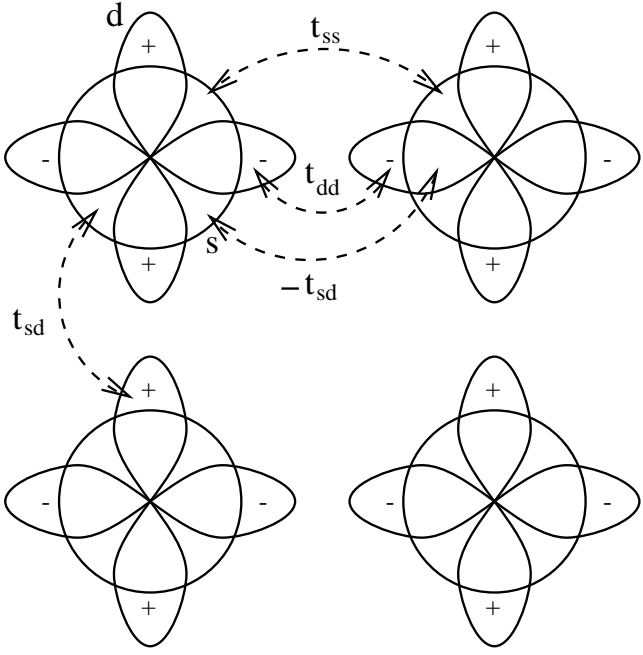}
 \end{center}
 \vspace{0mm}

 \caption{Sketch of the hopping terms
between the two orbitals in the unit cell.}
\label{hopping}
 \end{figure}

{\it The model.}
We study the simplest model with the features discussed above.
We postpone to the conclusions the analysis of possible
generalizations. Following the scheme in\cite{HM89}, we assume
an atom with two orbitals per site on a slightly distorted square
lattice. The distortion makes the lattice orthorhombic. We 
assume that this distortion is small, and we will treat its 
effects as a perturbation. Using the tetragonal symmetry of
the undistorted lattice, we take the lowest lying orbital
to be $d_{x^2 - y^2}$, and the second 
orbital to be $s$ or $d_{3z^2 - r^2}$.
We define $\Delta = \epsilon_s - \epsilon_{x^2-y^2}$ as the
difference between the energy of these levels. Without loss
of geenrality, we set $\epsilon_d = 0$. We assume that
there is only repulsion between electrons in the $d_{x^2 - y^2}$
orbital, $U$. Hopping can take place only between nearest neighbors,
with amplitudes $t_{ss} , t_{sd}$ and $t_{dd}$. Finally,
crystal field effects associated to 
the orthorhombic distortion induce an hybridization between
the $s$ and $d_{x^2 - y^2}$ in the same site, $V_{CF}$.
The hamiltonian is:
\begin{widetext}
\begin{eqnarray}
{\cal H} &= &{\cal H}_{ion} + {\cal H}_{tunn} \nonumber \\
{\cal H}_{ion} &= &\sum_{\sigma i} \Delta c^\dag_{s \sigma i}
c_{s \sigma i}
+ U n_{d \uparrow i} n_{d \downarrow i} + \sum_{\sigma i} V_{CF} 
c^\dag_{s \sigma i} c_{d \sigma i} + h. c. \nonumber \\
{\cal H}_{tunn} &= &\sum_{ij \sigma} t_{dd}  c^\dag_{d \sigma i}
c_{d \sigma j} + t_{ss} c^\dag_{s \sigma i} c_{s \sigma j}
\pm t_{sd} c^\dag_{s \sigma i} c_{d \sigma j} + h. c.
\label{hamil}
\end{eqnarray}
\end{widetext}
For simplicity, we neglect here possible differences between
the hoppings in the two directions due to the asymmetry of
the lattice\cite{NK01}, which is included through the crystal field potential
$V_{CF}$ only.
Note that, due to the different symmetries of the two orbitals,
the hopping between an $s$ and a $d$ orbital has opposite sign along
the two axes of the lattice (see fig.[\ref{hopping}]).

We now assume that $V_{CF} \ll U , | t_{ss} | , | t_{sd} | , 
| t_{dd} |
\ll \Delta$. The lowest lying eigenstates of ${\cal H}_{ion}$
for each occupancy are:
\begin{eqnarray}
| 0 \rangle &= & | 0 \rangle \nonumber \\
| 1 \rangle_\sigma &\approx &\left( c^\dag_{d \sigma} + \frac{V_{CF}}{
\Delta} c^\dag_{s \sigma} \right)  | 0 \rangle \nonumber \\
| 2 \rangle &\approx &c^\dag_{d \uparrow} c^\dag_{d \downarrow} | 0 \rangle +
\frac{V_{CF}}{\Delta - U} \left( c^\dag_{d \uparrow} c^\dag_{s \downarrow}
+ c^\dag_{s \uparrow} c^\dag_{d \downarrow} \right) | 0 \rangle
\label{states}
\end{eqnarray}
with energies:
\begin{eqnarray}
E_0 &= &0 \nonumber \\
E_1 &\approx &- \frac{V_{CF}^2}{\Delta} \nonumber \\
E_2 &\approx &U - 2 \frac{V_{CF}^2}{( \Delta - U )}
\label{levels}
\end{eqnarray}
Using eq.(\ref{states}) we define an effective hopping which depends
on the occupancy of the site. We find:
\begin{eqnarray}
\langle 0_i | \langle 1_{\sigma j} | {\cal H}_{tunn}
| 1_{\sigma i} \rangle | 0_j \rangle &\approx &t_{dd} \pm 2 t_{sd} 
\frac{V_{CF}}{\Delta} \nonumber \\
\langle 1_{\uparrow i} | \langle 1_{\downarrow j} |
{\cal H}_{tunn} | 2_i \rangle | 0_j \rangle &\approx &
t_{dd} \pm t_{sd} \left( \frac{V_{CF}}{\Delta}
+ \frac{V_{CF}}{\Delta - U} \right) \nonumber \\
\langle 1_{\sigma i} | \langle 2_j | {\cal H}_{tunn} |
| 2_i \rangle 1_{\sigma j} \rangle &\approx  &t_{dd} \pm 2 t_{sd}
\frac{V_{CF}}{\Delta - U}
\end{eqnarray} 
We can now write an effective hamiltonian:
\begin{widetext}
\begin{equation}
{\cal H}_{eff} =  \sum_i \tilde{U} \tilde{n}_{i \uparrow} \tilde{n}_{j \downarrow}
+ \sum_{\sigma ij} \tilde{t}_\pm \tilde{c}^\dag_{\sigma i}
\tilde{c}_{\sigma j} \pm \delta t ( \tilde{n}_{\sigma i} +
\tilde{n}_{\sigma j} ) \tilde{c}^\dag_{- \sigma i} 
\tilde{c}_{- \sigma j} + h. c.
\label{hamil_eff}
\end{equation}
\end{widetext}
where we have shifted the origin of energies 
by $- V_{CF}^2 / \Delta$, see eq.(\ref{levels}), and:
\begin{eqnarray}
\tilde{c}_{\sigma i} &\approx 
&c_{d \sigma i} + \frac{V_{CF}}{\Delta} c_{s \sigma i}
\nonumber \\
\tilde{t}_\pm &\approx &t_{dd} \pm 2 t_{sd} \frac{V_{CF}}{\Delta}
\nonumber \\
\delta \tilde{t} &\approx &2 t_{sd} \frac{V_{CF} U}{\Delta^2}
\nonumber \\
\tilde{U} &\approx &U - 2 \frac{U V_{CF}^2}{\Delta^2}
\end{eqnarray}
Note that, as we are expanding to first order in $V_{CF} / \Delta$,
we can neglect normalization terms in the definition of the 
electron operators $\tilde{c}_{\sigma i}$. For the same reason,
$\delta \tilde{t}$ has the same absolute value along
the two axes of the lattice. The symmetries of the orbitals involved
imply that, to first order, there are not
next nearest neighbor assisted hopping terms.

{\it Superconducting solution.}
The energy bands of the  effective hamiltonian, eq.(\ref{hamil_eff}),
are:
\begin{equation}
\epsilon_{k_x k_y} = \left( \tilde{t}_+ + \frac{n}{2}
\delta \tilde{t} \right) \cos ( k_x ) +
\left( \tilde{t}_- - \frac{n}{2} \delta \tilde{t} \right) \cos ( k_y )
\end{equation}
Where $n$ is the number of electrons in the unit cell.

The pairing interaction is\cite{HM89}:
\begin{equation}
V_{k_x k_y k_x' k_y'} = \tilde{U} + \delta \tilde{t}
\left[ \cos ( k_x ) - \cos ( k_y ) + \cos ( k_x' ) - \cos ( k_y ' )
\right]
\end{equation}
The superconducting gap must be of the form:
\begin{equation}
\Delta_{sc \, k_x k_y}  = a + b \left[ \cos ( k_x ) - \cos ( k_y ) \right]
\label{gap}
\end{equation}
Following ref.\cite{HM89}, at the transition temperature these
coefficients satisfy:
\begin{widetext}
\begin{eqnarray}
a &= &a \left[ \tilde{U} I_0 - \delta \tilde{t} ( I_x - I_y ) \right]
+ b \left[ \tilde{U} ( I_x - I_y ) + \delta \tilde{t}
( I_{xx} - 2 I_{xy} + I_{yy} ) \right] \nonumber \\
b &= &a \delta \tilde{t} I_0 - b \delta \tilde{t} ( I_x - I_y )
\label{coefficients}
\end{eqnarray}
\end{widetext}
where:
\begin{eqnarray}
I_0 &= &\int_{ k_x k_y}^{\epsilon_{k_x k_y} < \epsilon_F}
\frac{
n ( \epsilon_{k_x k_y} / T )}{\epsilon_{k_x k_y} - \epsilon_F} 
\nonumber \\
I_x &= &\int_{ k_x k_y}^{\epsilon_{k_x k_y} < \epsilon_F}
\frac{ \cos ( k_x )
n ( \epsilon_{k_x k_y} / T )}{\epsilon_{k_x k_y} - \epsilon_F}
\nonumber \\
I_{xx} &= &\int_{ k_x k_y}^{\epsilon_{k_x k_y} < \epsilon_F}
\frac{ \cos^2 ( k_x )
n ( \epsilon_{k_x k_y} / T )}{\epsilon_{k_x k_y} - \epsilon_F}
\label{integrals}
\end{eqnarray}
and similar expressions for $I_y , I_{xy}$ and $I_{yy}$.
The function $n ( \omega / T )$ is the Fermi-Dirac distribution,
and $\epsilon_F$ is the Fermi energy.

From eqs.(\ref{coefficients}), the critical temperature is
given by:
\begin{widetext}
\begin{equation}
0 = 1 + \tilde{U} I_0 - 2 \delta \tilde{t} ( I_x - I_y )
+ \delta \tilde{t}^2 \left[ ( I_x - I_y )^2 -
I_0 ( I_{xx} - 2 I_{xy} + I_{yy} ) \right]
\label{temperature}
\end{equation}
\end{widetext}
The integrals in eqs.(\ref{integrals}) diverge as $- \log [ 
 ( N ( \epsilon_F ) T ]$ at low
temperatures, where $N ( \epsilon_F )
\sim \tilde{t}^{-1}$ is the density of states at the Fermi level,
and $\tilde{t} = ( \tilde{t}_+ = \tilde{t}_- ) /2$.
As we are assumming a weak orthorhombic
distortion, $I_x \approx I_y$. Hence, eq.(\ref{temperature}) always
has a solution at low temperatures. The critical temperature
is given, approximately, by:
\begin{equation}
T_c \sim c_1 \tilde{t} e^{- ( c_2 \tilde{U} )
/ N ( \epsilon_F ) \delta \tilde{t}^2} \sim c_1' t
e^{- ( c_2 \Delta^4 ) / [ t^2 N ( \epsilon_F ) V_{CF}^2 U ]}
\label{tc}
\end{equation}
where we assume $t_{dd} \sim t_{sd} \sim t$. $c_1$ and $c_2$
are numerical constants.
The expression in eq.(\ref{tc})
implies that the superconductivity is supressed
as $\Delta \rightarrow \infty$. It is enhanced by the 
the lattice asymmetry, described by $V_{CF}$, and by the existence of a
finite $U$, as the assisted hopping term requires the presence
of electron-electron interactions. Near half filling, the van Hove
singularity in the density of states implies $T_c \propto t
e^{- c \sqrt{\lambda}}$, 
where $\lambda = \Delta^4 / ( t V_{CF}^2 U )$\cite{HS86,LB87,F87}.
The expression in eq.(\ref{tc}) ceases to be valid near the band edges,
as $I_{xx} -2 I_{xy} + I_{yy} \rightarrow 0$. It can be shown that,
in this limit:
\begin{equation}
\lim_{n \rightarrow 0} I_{xx} -2 I_{xy} + I_{yy} \propto n
\end{equation}
where $n$ is the number of carriers per unit cell. Then:
\begin{equation}
\lim_{n \rightarrow 0} T_c \sim c_1' t
e^{- ( c_2 \Delta^4 ) / [ n t V_{CF}^2 U ]} \rightarrow 0
\end{equation}

We can also calculate:
\begin{equation}
\frac{a}{b} \sim \frac{\delta \tilde{t}^3 N^2 ( \epsilon_F ) }
{U I_0} \sim \frac{t^3 N^2 ( \epsilon_F ) U^2 V_{CF}^3}{\Delta^6}
\end{equation}
so that $a/b \rightarrow 0$ as $\Delta$ becomes the largest energy
in the problem. In this limit, the gap, eq.(\ref{gap}), will
have $d_{x^2 - y^2}$ symmetry. 

It is interesting to compare the value in eq.(\ref{tc}) with the
critical temperature deduced from the Kohn-Luttinger analysis for
the Hubbard model without assisted hopping terms, $T_{c KL}$. The effective 
coupling constant arises from the screened interaction, and, to
lowest order in $U$, it goes as $U^2 N ( \epsilon_F )$.
Hence, $T_{c KL} \sim d_1  t e^{- d_2 / ( U N ( \epsilon_F ) ]^2}$. 
Thus, within the perturbative approach used here, we find
$T_c \gg T_{c KL}$, although $T_c$ also depends on the strength
of the orthorhombic distortion.

\begin{figure}
 \begin{center}
        \leavevmode
        \epsfxsize=85mm 
        \epsfbox{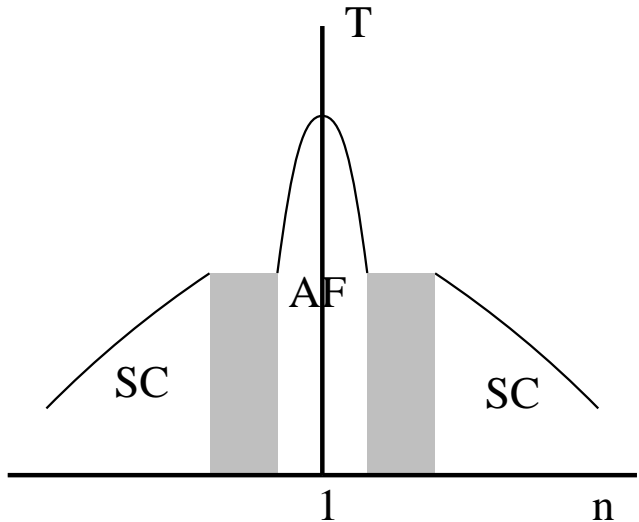}
 \end{center}
 \vspace{0mm}

 \caption{Sketch of expected phase diagram for the model
described by eq.(\protect{\ref{hamil}}),
see text for details. The present analysis
is insufficient to characterize the intermediate regions
between the antiferromagnetic and superconducting phases, shown
shaded in the figure. 
}
\label{dwave_phased}
\end{figure}

{\it Phase diagram.}
So far, we have only considered the superconducting instability.
Near half filling, it is well known that the hamiltonian in
eq.(\ref{hamil}) has nesting properties, and antiferromagnetism is
favored. The corresponding N\`eel temperature is proportional to
the gap, so that\cite{H85}:
\begin{equation}
T_N \sim b_1 t e^{- b_2 / \sqrt{U N ( \epsilon_F )}}
\end{equation}
Assuming that $V_{CF} \ll \Delta$, we find that, near half filling,
$T_c \ll T_N$, and the system will be antiferromagnetic. This phase
dissappears at densities, measured from half filling,
$n \sim e^{- b_2 / \sqrt{U N ( \epsilon_F )}}$. A sketch
of the expected phase diagram is shown in Fig.[\ref{dwave_phased}].
Th method used here cannot be used to
elucidate the nature of the transition from
the antiferromagnetic phase at half filling to the superconducting
phase. Hartree-Fock studies of the Hubbard model suggest that
an orthorhombic distortion favors the formation of stripes\cite{NK01}.
Alternatively, phase separation is also possible\cite{GGA00}.
Higher order terms in $V_{CF} / \Delta , U / \Delta$, not considered here,
will break the electron-hole symmetry shown in
Fig.[\ref{dwave_phased}]\cite{HM89,H93}.
It is interesting to note that the single band Hubbard model can
show an intrinsic tendency towards an orthorhombic distortion near half
filling\cite{HM00}.

{\it Conclusions.}
We have studied a variation of the assisted hopping model, which includes
the effects of other atomic levels in Hubbard hamiltonians\cite{HM89,H93}
(see also\cite{Letal99}).
We have avoided the difficulties associated to intermediate coupling 
situations by restricting the study to a well defined weak coupling case,
where standard perturbation theory is reliable (for a similar
approach to other assisted hopping problem, see\cite{G02}).

The induced assisted hopping terms lead to d-wave superconductivity.
The arguments used here show that the symmetry of the order parameter,
for models with different atomic orbitals, is 
associated to the relative symmetry of these orbitals.
When the most relevant orbitals are of the same symmetry, only
s-wave superconductivity is possible, in agreement with\cite{HM89,H93}.
It is interesting to note that the symmetry of the order parameter
does not arise from the anisotropies of the Fermi surface, as for
the Kohn-Luttinger mechanism. We have shown, that in the weak coupling
regime, the value of the critical temperature arising from
the assisted hopping term tends to be larger than that due to
the Kohn-Luttinger mechanism. It is plausible that, in the same
manner, the presence of assisted hopping terms will enhance
the tendency of the standard Hubbard model towards superconductivity
in the intermediate coupling regime.

Finally, the present results can be extended to other systems 
where assisted hopping terms are likely to arise. In quantum dots,
where the sign of these terms can be random\cite{G02}, they can
enhance the tendency towards local pairing.

{\it Acknowledgements.}
I am grateful to J. E. Hirsch, for sharing his insights in the nature of
assisted hopping effects, and for many other helpful discussions.
The hospitality of the Kavli Institute for Theoretical Physics,
where this work was done,
is gratefully acknowledged. The KITP is supported by NSF
through grant PHY99-07949. Additional funding comes from MCyT (Spain)
through grant PB96-0875.

\end{document}